\documentclass[10pt,conference]{IEEEtran}
\usepackage{blindtext,graphicx}
\usepackage{times}
\usepackage[a4paper, total={184mm,239mm}]{geometry}

\IEEEoverridecommandlockouts
\usepackage{cite}
\usepackage{amsmath,amssymb,amsfonts}
\usepackage{algorithmic}
\usepackage{textcomp}
\usepackage{setspace}
\def\smallerspacecaption{\vspace{-2mm}}

\usepackage{floatrow}
\usepackage[caption=false]{subfig}
\usepackage[table, dvipsnames]{xcolor} 
\usepackage{fancyhdr}
\usepackage{graphicx}
\usepackage{url}
\usepackage{verbatim}
\usepackage{multirow}
\usepackage{hhline}
\usepackage[us]{datetime}
\usepackage{paralist}
\usepackage{color}
\usepackage{soul}
\usepackage{enumitem}
\usepackage[implicit=false,hidelinks]{hyperref}
\usepackage{stfloats}
\usepackage[font=footnotesize,labelsep=period,justification=centering]{caption}
\usepackage[ruled, vlined, norelsize]{algorithm2e}
\usepackage{pifont}
\newcommand{\cmark}{\ding{51}}
\newcommand{\xmark}{\ding{55}}

\usepackage{soul}
\newcommand{\drop}[1]{\textcolor{red}{#1}}
\renewcommand{\drop}[1]{}
\definecolor{cadmiumgreen}{rgb}{0.0, 0.42, 0.24}

\newdimen\arrayruleHwidth
\makeatletter
\def\Hline{\noalign{\ifnum0=`}\fi\hrule \@height \arrayruleHwidth
\futurelet \@tempa\@xhline}
\makeatother

\makeatletter
\def\blfootnote{\xdef\@thefnmark{}\@footnotetext}
\makeatother

\shortdate
\settimeformat{ampmtime}

\makeatletter
\newcommand*\tabsize{%
	   \@setfontsize\tabsize{6}{7.2}%
}
\makeatother

\renewcommand{\arraystretch}{1.03}

\renewcommand{\IEEEbibitemsep}{0.3pt}
\makeatletter
\IEEEtriggercmd{\reset@font\normalfont\fontsize{6.8pt}{7.11pt}\selectfont}
\makeatother
\IEEEtriggeratref{1}

\makeatletter
\renewcommand\footnoterule{%
  \kern-3\p@
  \hrule\@width 0.5\columnwidth
  \kern2.6\p@}
  \makeatother

\begin{document}

\title{\huge GNNUnlock: \underline{G}raph \underline{N}eural \underline{N}etworks-based Oracle-less \underline{Unlock}ing Scheme for Provably Secure Logic Locking \vspace{-5pt}}

\author{Lilas~Alrahis$^\Upsilon$, Satwik~Patnaik$^\dag$, 
Faiq~Khalid$^\S$, Muhammad~Abdullah~Hanif$^\S$, Hani~Saleh$^\Upsilon$,\\ Muhammad~Shafique$^\ddag$, and Ozgur~Sinanoglu$^\ddag$\\[1ex]
$^\Upsilon$Department of Electrical Engineering and Computer Science, Khalifa University, Abu Dhabi, UAE\\ 
$^\dag$Electrical \& Computer Engineering, Texas A\&M University, College Station, Texas, USA\\
$^\S$Institute of Computer Engineering, Technische Universität Wien, Vienna, Austria\\
$^\ddag$Division of Engineering, New York University Abu Dhabi, UAE\\
\normalsize{lilas.alrahis@ku.ac.ae}, \normalsize{satwik.patnaik@tamu.edu},
\normalsize{\{muhammad.shafique, ozgursin\}@nyu.edu}
}
\maketitle

\renewcommand{\headrulewidth}{0.0pt}
\thispagestyle{fancy}
\pagestyle{fancy}
\cfoot{
\copyright~2021 IEEE.
This is the author's version of the work.
It is posted here for personal use.
	Not for redistribution.\\
	The definitive Version of Record will be published in
	Proc.\ Design, Automation and Test in Europe (DATE) 2021\\
}

\begin{abstract}
Logic locking is a holistic design-for-trust technique that aims to protect the design intellectual property (IP) from untrustworthy entities throughout the supply chain.
Functional and structural analysis-based attacks successfully circumvent state-of-the-art, provably secure logic locking (PSLL) techniques. However, such attacks are not holistic and target specific implementations of PSLL. 
Automating the detection and subsequent removal of protection logic added by PSLL while accounting for \textit{all} possible variations is an open research problem.

In this paper, we propose GNNUnlock, the first-of-its-kind oracle-less machine learning-based attack on PSLL \textit{that can identify any desired protection logic without focusing on a specific syntactic topology}. 
The key is to leverage a well-trained graph neural network (GNN) to identify all the gates in a given locked netlist that belong to the targeted protection logic, \textit{without requiring an oracle}. 
This approach fits perfectly with the targeted problem since a circuit is a graph with an inherent structure and the protection logic is a sub-graph of nodes (gates) with specific and common characteristics. 
GNNs are powerful in capturing the nodes' neighborhood properties, facilitating the detection of the protection logic.
To rectify any misclassifications induced by the GNN, we additionally propose a connectivity analysis-based post-processing algorithm to successfully remove the predicted protection logic, thereby retrieving the original design. 

Our extensive experimental evaluation demonstrates that GNNUnlock is $99.24\%-100\%$ successful in breaking various benchmarks locked using stripped-functionality logic locking~\cite{yasin_CCS_2017}, tenacious and traceless logic locking~\cite{yasin_glsvlsi_2017}, and Anti-SAT~\cite{xie2016mitigating}. 
Our proposed post-processing enhances the detection accuracy, reaching $100\%$ for all of our tested locked benchmarks.
Analysis of the results corroborates that GNNUnlock is powerful enough to break the considered schemes under different parameters, synthesis settings, and technology nodes.
The evaluation further shows that GNNUnlock successfully breaks corner cases where even the most advanced state-of-the-art attacks~\cite{yang2019stripped,sirone2020functional} fail. 
We also open source our attack framework~\cite{webinterface}.
\end{abstract}

\begin{IEEEkeywords}
Logic locking,
IP protection,
Graph Neural Networks,
Machine Learning,
Oracle-less attack.
\end{IEEEkeywords}

\section{Introduction}
\label{sec:Introduction}

Logic locking (LL) is a design-for-trust technique that aims to protect the design intellectual property throughout the supply chain. 
The design's functionality is locked with a secret key (driven from an on-chip tamper-proof memory) and a set of added key-gates. 
The locked design functions correctly only when the correct secret key is in place. 
Earlier works in LL focused on the placement of key-gates and high output corruptibility~\cite{epic_journal,JV-Tcomp-2013,dupuis2014novel}.
However, these schemes were deemed vulnerable by the Boolean satisfiability (SAT)-based attack~\cite{Subramanyan_host_2015}.

Provably secure logic locking (PSLL) ensures an exponential number of SAT calls for a successful key recovery~\cite{xie2016mitigating,yasin_glsvlsi_2017,yasin_CCS_2017,mengli_provably_secure_camo,shakya2020cas}. 
In principle, a \textit{perturb} unit injects errors into the design for some protected input patterns. 
These errors are canceled out by a \textit{restore} unit when the correct key is applied. 
Recently, structural/functional \textit{oracle-less} attacks have emerged to circumvent PSLL~\cite{yasin_2017_sps,alrahis2019functional,yang2019stripped,sirone2020functional}.\footnote{\textit{Oracle-guided} attacks are successful in approximately circumventing specific instances of PSLL~\cite{shen2017double,shamsi2017appsat,xiaolin_bypass_attack_ches17}.
However, these attacks require a working chip for functional verification. 
We focus on the \textit{oracle-less} attacks as they pose a more significant and realistic threat, as highlighted by the LL community.} \textbf{Each attack aims to break a specific locking scheme, under restricted parameter settings and circuit formats}, resulting in a large set of attack strategies for different PSLL implementations. 
Developing a holistic attack on PSLL is an open research problem.
Table~\ref{tab:comparison} outlines the drawbacks of oracle-less attacks on PSLL.

\begin{table}[tb]
\centering
\scriptsize
\smallerspacecaption
\caption{\textsc{Capabilities offered by oracle-less attacks}}
\label{tab:comparison}
\resizebox{\textwidth}{!}{%
\setlength\tabcolsep{1.9pt}
\renewcommand\arraystretch{0.9}
\begin{tabular}{cccc}
\hline
\textbf{Attacks} & \textbf{\begin{tabular}[c]{@{}c@{}}Different\\ Circuit Formats\end{tabular}} & \textbf{\begin{tabular}[c]{@{}c@{}}Different\\ Locking Schemes\end{tabular}} & \textbf{\begin{tabular}[c]{@{}c@{}}Different\\ Parameter Settings\end{tabular}} \\\hline
SPS~\cite{yasin_2017_sps} & \color{red}{\xmark} & \color{red}{\xmark} & \color{cadmiumgreen}{\cmark}
\\ \hline
RE-based~\cite{alrahis2019functional} & \color{red}{\xmark} & \color{red}{\xmark} & \color{cadmiumgreen}{\cmark} 
\\ \hline
FALL~\cite{sirone2020functional} & \color{red}{\xmark} & \color{red}{\xmark} & \color{red}{\xmark} 
\\ \hline
SFLL-HD-Unlocked~\cite{yang2019stripped} & 
\color{red}{\xmark} & \color{red}{\xmark} & \color{red}{\xmark} 
\\ \hline
\textbf{GNNUnlock} & 
\color{cadmiumgreen}{\cmark} & 
\color{cadmiumgreen}{\cmark} & 
\color{cadmiumgreen}{\cmark} 
\\ \hline
\end{tabular}
}
\end{table}

\subsection{Motivation and Research Challenges}

\textbf{Restricted Parameter Settings:} Stripped functionality logic locking (\textit{SFLL-HD$^h$})~\cite{yasin_CCS_2017} showed resilience against all known attacks on LL.
SFLL-HD$^h$ protects all input patterns from a Hamming distance $h$ from the secret key $K$. 
Sirone \textit{et al.}~\cite{sirone2020functional} developed functional analysis attacks (FALL) on SFLL-HD$^h$ which retrieved the secret key \textit{without} requiring an oracle. 
The attack algorithms are based on a set of derived Lemmas defining functional properties of the protection logic used in SFLL-HD$^h$. 
The derived Lemmas hold for specific ranges of $h$ values. 
Hence, by definition, the attack algorithms cannot break all cases of SFLL-HD$^h$. 
For example, their \textit{AnalyzeUnateness} algorithm is only applicable when $h=0$, and their \textit{Hamming2D} algorithm is only applicable when $h \leq K/4$. 
Their third algorithm \textit{SlidingWindow} should apply to larger $h$ values in principle, but it does not fare well since it requires computationally hard SAT calls. 
Yang \textit{et al.}~\cite{yang2019stripped} proposed the \textit{SFLL-HD-Unlocked} attack that performs connectivity analysis on the locked circuit followed by a Gaussian-elimination-based analysis to recover the secret key \textit{without} requiring an oracle. 
Because of the use of Gaussian-elimination, the attack does not work when $h\leq4$ due to the composition of singular matrices.

To demonstrate the shortcomings of existing attacks, we lock selected ISCAS-85 and ITC-99 benchmarks using SFLL-HD$^h$ with $K/h=2$, multiple times, generating \texttt{192} locked designs. 
\textit{Such a locking ratio is critical as it achieves the highest resilience to removal attacks, as indicated in~\cite{yasin_CCS_2017}}. 
When the FALL attacks~\cite{sirone2020functional} were launched on these locked benchmarks; they reported \texttt{0} keys, failing to break any of the designs. 
Similarly, SFLL-HD-Unlocked attack~\cite{yang2019stripped} also was unable to recover the key for any design. 
\textit{This analysis highlights the need for a holistic approach that can break SFLL-HD$^h$ and other PSLL schemes under all the possible scenarios}.

\textbf{Restricted Circuit Formats:} The prior attacks on PSLL accept restricted and non-standard circuit formats. 
For example, the work in~\cite{alrahis2019functional} showcased that functional reverse engineering (RE) aids in the detection and subsequent removal of the protection logic added by SFLL-HD$^h$. 
However, the approach is not generic and will only work on designs synthesized using only 2-input gates. 
This attack achieves low accuracy results when launched on benchmarks synthesized with a standard cell library. 
For example, the \textit{restore} unit detection percentage in the ISCAS-85 benchmark \texttt{c7552} locked with $K=32$ and $h=0$ drops from $100\%$ to $0\%$. Moreover, both attacks~\cite{yang2019stripped,sirone2020functional} accept circuits in bench format (the latter~\cite{sirone2020functional} requires topologically sorted AND-OR-INVERT gates), which is a non-industry format, as opposed to \textit{Verilog/VHDL}. 
Thus, these techniques cannot be employed in the real-world design flow.

\textbf{Locking Scheme-specific Attacks:} All the aforementioned oracle-less attacks are locking scheme-specific. 
They target a distinct protection implementation (SFLL-HD$^h$ in this case) and are not scalable to other variants (such as \textit{Anti-SAT}~\cite{xie2016mitigating}). 
The signal probability skew (SPS) attack~\cite{yasin_2017_sps} is another oracle-less attack targeting Anti-SAT.
The SPS attack looks for two oppositely skewed nets feeding an AND gate, a property specific to the Anti-SAT protection and not to other PSLL schemes, and hence it is another scheme-specific attack.

\textbf{Associated Research Challenges:} Automating the detection of the protection logic added by PSLL while accounting for all possible variations among the implementations is an open research problem posing the following important challenges.

\begin{enumerate}[leftmargin=*]

\item \textit{Recognizing all implementation variants:} The protection logic structure depends on (i)~the key-value, (ii)~the Hamming distance value (for the case of SFLL-HD), (iii)~the choice of the logic blocks (for the case of Anti-SAT), and (iv)~the key-size. Hence, different settings will result in distinct topologies. 
Thus, utilizing an exact matching algorithm would be a naive approach. 

\item \textit{Handling different synthesis settings and circuit formats:} The structure of the protection logic and its integration with the original design depends on the synthesis procedures and the target technology library. 
The attack model should be able to understand and process different circuit formats without having to modify the attack tactic.

\end{enumerate}

\subsection{Our Novel Concept and Contributions}

To address the above challenges, we propose the \textbf{\textit{GNNUnlock} framework} that identifies all variants of the targeted protection logic. 
It is the first-ever concept to leverage graph neural networks (GNNs) along with node connectivity analysis to identify all the gates in a given locked netlist that form the protection logic, as depicted in Fig.~\ref{fig:GNN_attack}. 
This work is inspired by the ability of advanced machine learning (ML) models to adapt to new unseen data and the fact that a circuit is a graph with an inherent structure to it. 
Unlike other ML models, GNNs leverage the graph structure and node (gate) features to learn a representation for each node capturing its neighborhood characteristics.\footnote{We use the terms interchangeably gate or node, and circuit or graph.}
GNNs are a fitting choice for identifying the protection logic, which can be conceived as a sub-graph with specific properties.
The GNN learns the trend of the protection logic and not a syntactic implementation of it, and hence GNNUnlock deals with variations naturally. 
Our GNNUnlock framework employs the following techniques:

\begin{figure}[tb]
\centering
\includegraphics[width=0.9\textwidth]{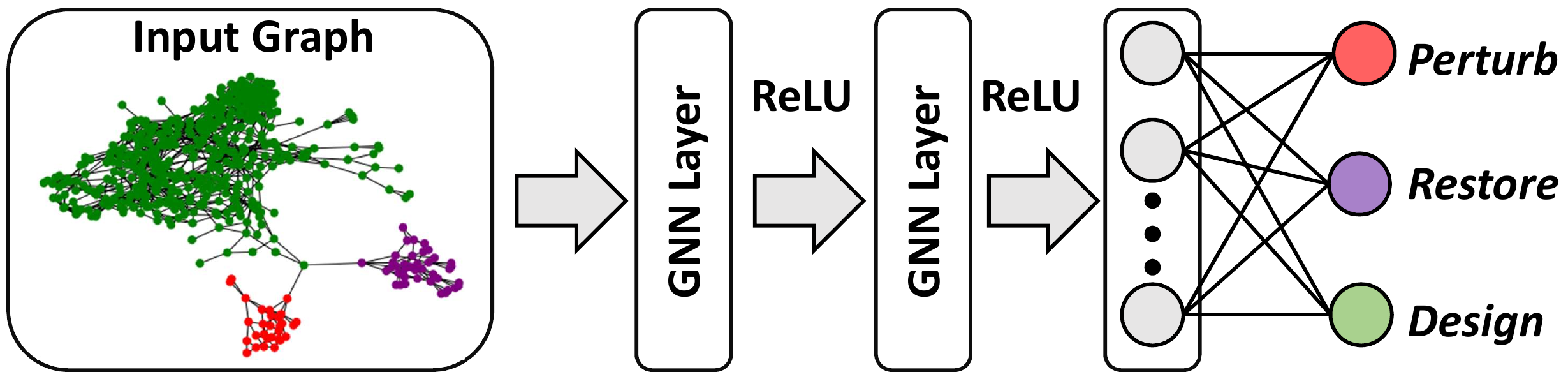}
\smallerspacecaption
\caption{High-level view of our work: Breaking PSLL using GNN.\vspace{-5pt}}
\label{fig:GNN_attack}
\end{figure}

\begin{enumerate}[leftmargin=*]

\item A framework for the \textbf{netlist-to-graph transformation (Section~\ref{Sec:transfromation})} is developed which implements automatic \& efficient feature extraction to capture each gate's functionality and connectivity in the gate-level netlist (in any format). 

\item \textbf{GNN learning on locked circuits is achieved (Section~\ref{Sec:attack_gnn})}, allowing the GNN to identify the structural features of all the nodes in the targeted protection logic.

\item \textbf{Post-processing rectification procedure (Section~\ref{Sec:post_processing})} guided by the connectivity of the circuit and by the predictions of the GNN (on the under-attack circuit) is developed to remove the identified protection logic effectively.

\end{enumerate}

The effectiveness of GNNUnlock is showcased by breaking \texttt{564} benchmarks locked using three different PSLL techniques -- SFLL-HD~\cite{yasin_CCS_2017}, \textit{TTLock}~\cite{yasin_glsvlsi_2017}, and Anti-SAT~\cite{xie2016mitigating}.
The effect of different $K$, different $h$ values (for the case of SFLL-HD, and other technology libraries are considered when evaluating GNNUnlock comprehensively. 
All the files locked at the register-transfer level (RTL) are synthesized following the commercial ASIC-design flow using Synopsys tools.
\textbf{We also make our attack framework available online at~\cite{webinterface}.}
\section{Background and Related works}
\label{sec:backgroud_RW}

\subsection{Provably Secure Logic Locking (PSLL)} 
\subsubsection{Anti-SAT~\cite{xie2016mitigating}} 

In the Anti-SAT block, the outputs of two complementary logic functions ($gl1$ and $\overline{gl2}$), locked with two sets of keys are fed into an AND gate, as shown in Fig.~\ref{fig:SFLL}a. 
When the correct key is in place the output of the AND gate $Y$ is \texttt{0}. For a wrong key-input, the output signal $Y$ could be \texttt{0} or \texttt{1} depending on the input $X$. 
$Y$ is XORed with an internal net in the original netlist, locking it. 
Having the $Y$ signal highly skewed to \texttt{0} ensures resilience against the SAT-based attack. 
Nevertheless, Anti-SAT is vulnerable to structural analysis-based removal attacks~\cite{yasin_2017_sps}.

\subsubsection{TTLock~\cite{yasin_glsvlsi_2017} and SFLL-HD~\cite{yasin_CCS_2017}} 
\begin{figure*}[tb]
\centering
\includegraphics[width=0.95\textwidth]{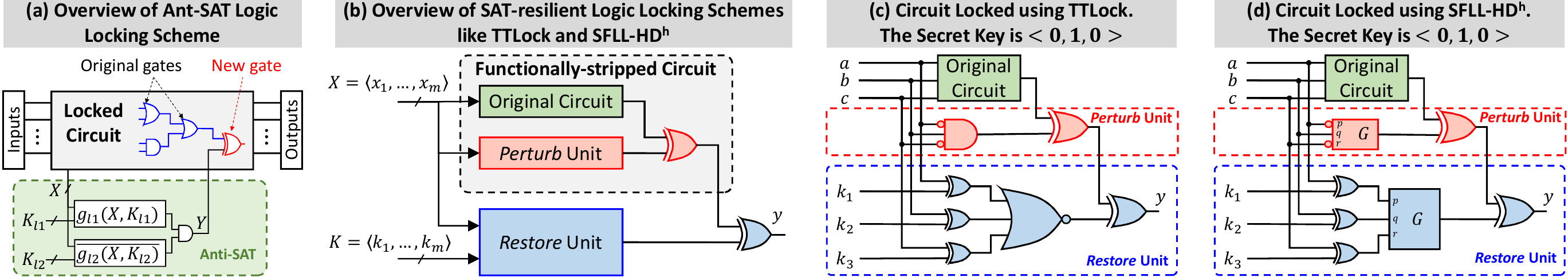}
\smallerspacecaption
\caption{Logic locking using Anti-SAT~\cite{xie2016mitigating}, TTLock~\cite{yasin_glsvlsi_2017}, and SFLL-HD~\cite{yasin_CCS_2017}. $G$ block in (d) performs Hamming distance checking using adders and comparators. \vspace{-5pt}}
\label{fig:SFLL}
\end{figure*}

modify the original design, as shown in Fig.~\ref{fig:SFLL}b, to achieve resilience against removal attacks.
TTLock protects an input pattern that is the same as the locking key. Tracing the key-inputs (KIs) helps identify the \textit{restore} logic of TTLock, which is a basic comparator. 
However, the \textit{perturb} unit structure depends on the selected secret key, and not controlled by the external KIs, as demonstrated in Fig.~\ref{fig:SFLL}c. 
SFLL-HD$^h$ protects all input patterns that are an $h$ Hamming distance away from the secret key. 
TTLock is equivalent to SFLL-HD$^0$. For SFLL-HD$^{h>0}$, the \textit{restore} and \textit{perturb} units are Hamming distance checkers ($G$), as illustrated in Fig.~\ref{fig:SFLL}d. 
The structure of the \textit{perturb} unit depends on the hard-coded key, making it difficult to be traced.

\vspace{-0.5ex}
\subsection{Graph Neural Network (GNN)} 

In GNNs, nodes aggregate information from their neighbors using neural networks. 
Each node gathers information from its neighbors in a previous layer, where each one of these neighbors collects information from the previous layer. 
The aggregation functions in \textit{GraphSAGE}~\cite{hamilton2017inductive} approach learn to aggregate feature information to obtain embeddings for the nodes, which are then used to perform the desired task such as node classification. 
The same aggregation parameters are shared for all nodes. 
Hence, once the model is trained, those parameters can generate embeddings on entirely unseen graphs. 
We follow such an inductive approach since we train on a set of locked benchmarks and \textit{test the model on unseen benchmarks locked with unknown key-values}. 
We adopt the GraphSAGE architecture in which each node's previous state is concatenated with its neighbors' current state. The mean aggregation function is used to update the state of each node. 

Due to neighborhood aggregation, the deeper the GNN model is, the more multi-hop neighbors get incorporated for the root node's computation. 
Hence, training time could grow exponentially with GNN depth~\cite{zeng2019graphsaint}. 
Several methods, including GraphSAGE, perform layer sampling to ensure a limited set of neighbors is considered for a node in the next layer. 
Such sampling speeds up training but suffers from scalability issues.
We leverage the graph sampling method \textit{GraphSAINT}~\cite{zeng2019graphsaint} to construct mini-batches for training. 
Rather than building a GNN on full training graph and then perform layer sampling, GraphSAINT samples the training graph and then builds a GNN on the sampled graph for each mini-batch,\footnote{In each mini-batch, forward and backward propagation is performed iteratively on the sampled GNN to update weights via stochastic gradient descent.} ensuring scalability concerning graph size and GNN depth. 
Their random walk-based sampler is used in our experiments.
\section{Attacker Model}
\label{sec:threat_model}

We assume that the attacker will be present in the untrusted foundry with access to the GDSII mask information and reverse engineering tools that facilitate the gate-level netlist extraction.
Besides, the attacker is aware of the locking algorithm and can distinguish between the regular primary inputs (PIs) and KIs. 
The attacker also knows the usage of the technology library and the usage of different synthesis settings. 
We do not assume the attacker to have access to an unlocked chip (i.e., an \textit{oracle-less} setting). 
For SFLL-HD$^h$~\cite{yasin_CCS_2017}, the attacker knows the Hamming distance value, while the type of logic function ($gl$) is known for the case of Anti-SAT~\cite{xie2016mitigating}. 
Note, all these assumptions are in line with the \textit{Kerckhoffs's principle}, which states that everything about the system should be known to an attacker \textit{except} for the value of the secret key.
\section{Proposed GNNUnlock Framework}
\label{sec:Proposed_attack}

\begin{figure*}[tb]
\centering
\includegraphics[width=0.95\textwidth]{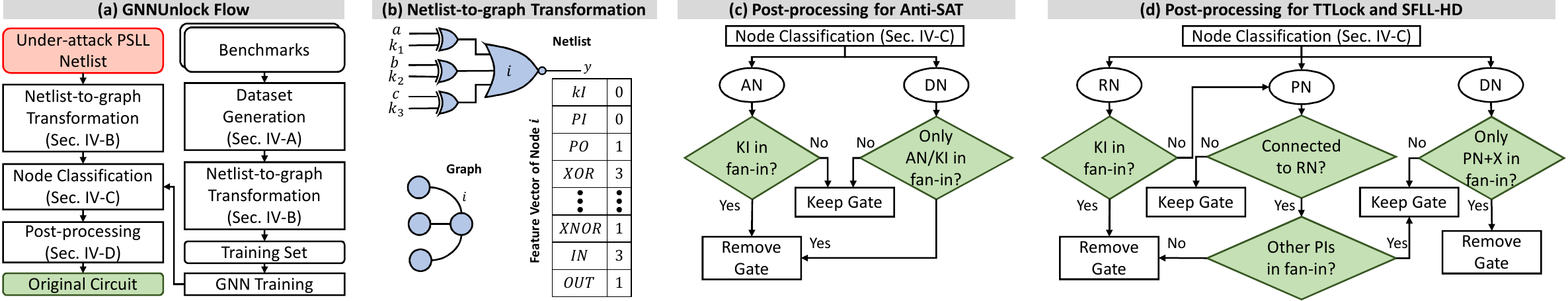}
\smallerspacecaption
\caption{Proposed GNNUnlock methodology. 
The \textit{Anti-SAT}, \textit{restore}, \textit{perturb}, and \textit{design} nodes are indicated by AN, RN, PN, and DN, respectively.
\vspace{-1pt}
}
\label{fig:GNNUnlock_flow}
\end{figure*}

Fig.~\ref{fig:GNNUnlock_flow} shows an overview of our methodology, with key steps discussed in the following subsections.

\subsection{Dataset Generation} 

Using the underlying locking scheme, we generate a set of locked benchmarks for training and validation. 
Each benchmark is locked several times with randomly generated key-values and different key-sizes $K$, which results in an extensive training set. 
The training process is not aware of the correct key-values. 
However, different key-values cause the generation of varying \textit{perturb} and \textit{Anti-SAT} structures, and hence variants are accounted for during training. GNNUnlock transforms locked benchmarks into graphs (Section~\ref{Sec:transfromation}) and labels the nodes. 
For the case of SFLL-HD$^{h>0}$/TTLock, a node belongs to the \textit{perturb}, the \textit{restore}, or the \textit{design} logic. For the case of Anti-SAT, a node belongs to the \textit{Anti-SAT} block or the \textit{design} logic.

GNNUnlock attacks each design independently by excluding its corresponding graphs from training/validation. For example, when attacking \texttt{b17\_C} from ITC-99, only the graphs of \texttt{b14\_C}, \texttt{b15\_C}, \texttt{b20\_C}, and \texttt{b21\_C} are used in training, while the graphs of \texttt{b22\_C} are used for validation to evaluate the model on unseen data, thereby avoiding biasing.

\subsection{Netlist-to-Graph Transformation}
\label{Sec:transfromation}

We model connectivity between the gates in a design using an undirected graph $G(I, J)$. 
The set $I$ of nodes represents all the gates, while the set $J$ of edges represents the wires. 
An example of a logic locked netlist and the corresponding graph is shown in Fig.~\ref{fig:GNNUnlock_flow}b. 
The set $I$ does not contain PIs, KIs, or primary outputs (POs) of the design. 
Each node in the graph is associated with a feature vector $\hat{f}$ that contains information describing the node's important characteristics, such as its in-degree ($IN$) and out-degree ($OUT$). 
It also contains information on whether the node is connected to a PI, a PO, or a KI. 
Additionally, $\hat{f}$ captures the type and number of gates appearing in the neighborhood of the node (all nodes within two-hops away). 
For example, $\hat{f_{i}}$ shows that gate $i$ is connected to a PO with $IN=3$ and $OUT=1$ (Fig.~\ref{fig:GNNUnlock_flow}b). 
It also indicates that three XOR gates and one XNOR gate exist in $i$'s neighborhood. 
The length of the feature vector $|\hat{f}|$ depends on the number of logic gates in the target library. 
Each locked design is represented by a graph instance with a corresponding adjacency matrix.
To feed multiple graphs of different sizes to the GNN, a block-diagonal matrix is created for each dataset. 
Each block represents the adjacency of one locked design.

\vspace{-1ex}
\subsection{GNN Topology}
\label{Sec:attack_gnn}

In all experiments reported in Section~\ref{sec:results}, graph sampling using the GraphSAINT method is performed.
A two-layer GNN is constructed using GraphSAGE architecture, with mean and concatenation aggregation, and the \textit{ReLU} activation function. Training stops after a fixed number of epochs based on convergence. 
The model with the best performance on the validation set is used to evaluate the test set accuracy. 
Details regarding the GNN model and sampling are shown in Table~\ref{tab:GNN_info}.

\subsection{Post-processing}
\label{Sec:post_processing}

Although our trained GNN achieves high node classification accuracy on its own (with an average of $99.99\%$, $99.95\%$, and $99.92\%$, for the case of Anti-SAT, TTLock, and SFLL-HD$^2$, respectively), we propose a post-processing algorithm to rectify any potential misclassifications to enhance the accuracy further.
The algorithm considers predictions made by the GNN, connectivity of the circuit, and known properties of the protection logic, which is explained further in the next subsections.

\subsubsection{Anti-SAT} Every node in the \textit{Anti-SAT} block must have at least one KI in its fan-in cone. 
If a node without KIs in its fan-in cone is predicted as an \textit{Anti-SAT} node, then the prediction will be dropped. 
For each \textit{predicted design} node, we check its fan-in cone. If there are only other \textit{predicted Anti-SAT} nodes in the fan-in cone, then the node initially classified as a \textit{design} node is considered part of the Anti-SAT block. The flow of our post-processing algorithm is presented in Fig.~\ref{fig:GNNUnlock_flow}c. 

\subsubsection{TTLock and SFLL-HD$^{h>0}$} Our post-processing algorithm for the case of SFLL-HD$^{h>0}$ and TTLock is presented in Fig.~\ref{fig:GNNUnlock_flow}d. 
The \textit{predicted restore} nodes are visited to identify possible protected inputs (set $X$). The following properties then guide the post-processing procedure. (i)~All \textit{restore} nodes have KIs in their fan-in cone. 
(ii)~All \textit{perturb} nodes have connections with the \textit{restore} nodes in the netlist and are controlled solely by protected inputs $X$. 
If a \textit{predicted restore} node has KIs in its fan-in cone, then the prediction is confirmed. 
Otherwise, the algorithm checks if the node is a \textit{perturb} node. 
A \textit{predicted perturb} node is verified if it has a connection to other \textit{predicted restore} nodes in the netlist and if the node has $X$ (and no other PIs) in its fan-in cone. 
In order not to misclassify \textit{perturb} nodes as \textit{design} nodes, the algorithm checks the fan-in cone of the predicted \textit{design} nodes. 
If a predicted \textit{design} node has $X$ and other predicted \textit{perturb} nodes in its fan-in, it is a \textit{perturb} node.

\begin{table}[tb]
\centering
\smallerspacecaption
\caption{\textsc{GNN configuration and sampling details. The \#classes for SFLL-HD$^{h>0}$ and TTLock is \texttt{3}, while for Anti-SAT it is \texttt{2}}}
\label{tab:GNN_info}
\resizebox{\textwidth}{!}{%
\renewcommand\arraystretch{0.9}
\begin{tabular}{cccc}
\hline
\multicolumn{2}{c}{\textbf{Architecture}} & \multicolumn{2}{c}{\textbf{Training and Sampling}} \\ \hline
\textbf{Input Layer} & {[}$|\hat{f}|$,$512${]} & \textbf{Optimizer} & Adam \\ \hline
\textbf{Hidden Layer 1} & {[}$1024$,$512${]} & \textbf{Learning Rate} & $0.01$ \\ \hline
\textbf{Hidden Layer 2} & {[}$1024$,$512${]} & \textbf{Dropout} & $0.1$ \\ \hline
\textbf{Output Layer} & {[}$512$,\#classes{]} & \textbf{Sampler} & Random Walk \\ \hline
\textbf{Aggregation} & Mean with concatenation & \textbf{Walk Length} & $2$ \\ \hline
\textbf{Activation} & ReLU & \textbf{Root Nodes} & $3000$ \\ \hline
\textbf{Classification} & Softmax & \textbf{Max \# Epochs} & $2000$ \\ \hline
\end{tabular}%
}
\end{table}
\section{Experimental Investigations}
\label{sec:results}

\subsection{Evaluation Setup, Tool Flow, and Evaluation Metrics} 

\begin{figure}[tb]
\centering
\includegraphics[width=0.97\textwidth]{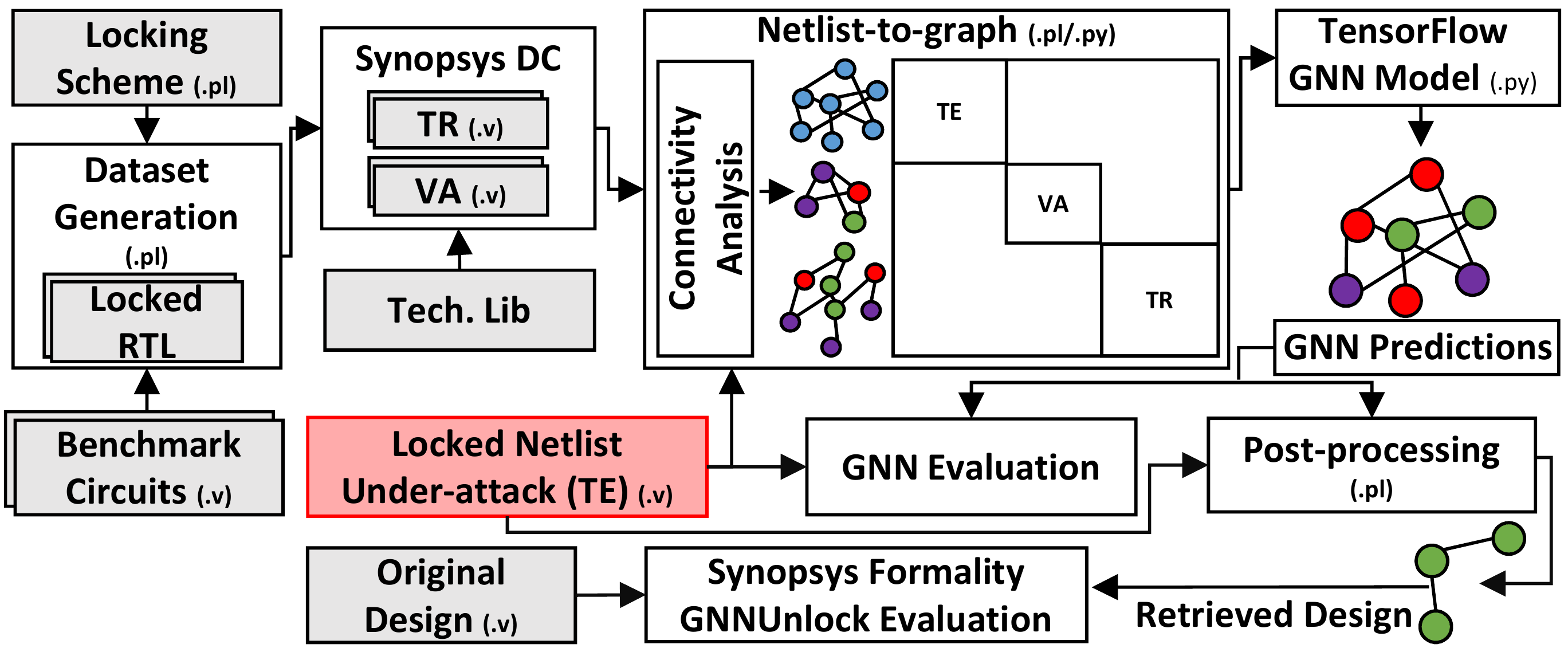}
\smallerspacecaption
\caption{Evaluation setup showing different components and platforms. TE, TR, and VA stand for testing, training, and validation, respectively.\vspace{-5pt}}
\label{fig:GNNUnlock_exp}
\end{figure}

GNNUnlock is evaluated on selected ISCAS-85 and ITC-99 benchmarks. 
Netlist-to-graph transformation is implemented in Perl/Python3. 
Tensorflow with Python3 implementation of GraphSAINT is used for GNN training~\cite{zeng2019graphsaint}. 
Training is performed on a single node with 24 cores (2x Intel Xeon CPUs E5-2695 v2@2.4 GHz), 256GB RAM, and one NVIDIA Tesla K20m GPU (2,496 CUDA cores and 5GB of GDDR5 memory). Fig.~\ref{fig:GNNUnlock_exp} summarizes the experimental setup.

\subsubsection{Dataset Generation for Anti-SAT}

The benchmarks (in bench format) were locked using the Anti-SAT locking binary provided by the authors.
Each ISCAS-85 benchmark is locked twice with $K:\{8, 16, 32, 64\}$, except for \texttt{c3540}, where $K=64$ is not considered due to the limited number of PIs in the design. 
In total, \texttt{30} Anti-SAT locked ISCAS-85 benchmarks are obtained. 
$K=8$ is used to evaluate the effectiveness of GNNUnlock in isolating the Anti-SAT block when its size is very small compared to that of the original design.
Each ITC-99 benchmark is locked twice with $K:\{32, 64, 128\}$ resulting in a total of \texttt{36} Anti-SAT locked ITC-99 benchmarks. Two labeled datasets for Anti-SAT block identification have been created, one for locked ISCAS-85 benchmarks and one for locked ITC-99 benchmarks. A feature vector $|\hat{f}|=13$ is associated with each node. 
\textit{The Anti-SAT locking binary only accepts circuits in bench format, which includes a restricted set of logic gates (\texttt{8} gates)}. 
Hence, only \texttt{8} out of the \texttt{13} features are required to represent the neighborhood of each node.
Rest of the features represent $IN$, $OUT$, and the connectivity to PIs, KIs, or POs.
\subsubsection{Dataset Generation for TTLock and SFLL-HD$^{h>0}$}
SFLL-HD$^h$ is implemented in Perl as described in~\cite{yasin_CCS_2017} to lock the benchmarks at RTL -- TTLock is the case when $h=0$. Each ISCAS-85 benchmark is locked thrice with $K:\{8, 16, 32, 64\}$ for $h:\{0,2\}$, except for the \texttt{c3540} benchmark where $K=64$ is not considered due to the limited number of PIs in the design. 
In total, \texttt{45} locked ISCAS-85 benchmarks for each $h$ are obtained. 
Each ITC-99 benchmark is locked thrice with $K:\{32, 64, 128\}$ for $h:\{0,2,4\}$, resulting in a total of \texttt{54} locked ITC-99 benchmarks for each setting of $h$. 
Locked RTL files are synthesized using the standard ASIC design flow for the $65nm$ LPe technology. 
Synthesis is performed using Synopsys Design Compiler. 
$|\hat{f}|=34$ is associated with each node. Since a full standard cell library is used for synthesis, a larger $|\hat{f}|$ (than the case of Anti-SAT) is required to capture all the possible combinational logic gates.

To verify that GNNUnlock handles different circuit formats and technologies, the Nangate $45nm$ open cell library is also used for synthesis when $h=2$. Varying the circuits' format affects $|\hat{f}|$ only. For this case, $|\hat{f}|=18$ is used. 
We also consider other corner cases for when $h:\{16, 32, 64\}$ with corresponding $K:\{32,64,128\}$, more details are in Section~\ref{sec:corner_cases}. The characteristics of all the datasets are listed in Table~\ref{tab:Datasets}.

\subsubsection{Evaluation Methods and Metrics}

GNNUnlock attacks each design independently, by excluding its corresponding graphs from training and validation. 
For example, when attacking the TTLock-ed \texttt{b17\_C} design from ITC, only the graphs of \texttt{b14\_C}, \texttt{b15\_C}, \texttt{b20\_C}, and \texttt{b21\_C} are included in the training set, resulting in \texttt{161,942} training nodes (\texttt{36} graphs), while the graphs of \texttt{b22\_C} are included only in the validation set (to evaluate the model on unseen data), resulting in \texttt{74,753} validation nodes (\texttt{9} graphs). And only the graphs of \texttt{b17\_C} are tested on, resulting in \texttt{110,685} testing nodes (\texttt{9} graphs).

The predictions of the GNN are compared with the true labels of the nodes to evaluate the performance of the attack. 
We report the accuracy and the non-averaged precision, recall, and F1-score for each classifier. 
We further evaluate GNNUnlock by removing the predicted protection logic to retrieve the unlocked design without accessing the true labels (Section~\ref{Sec:post_processing}). 
The recovered benchmark is then compared with the original benchmark via circuit-equivalence using Synopsys Formality.

\subsection{Breaking Anti-SAT with GNNUnlock}
\label{sec:breaking_antisat}

For the ISCAS-85 dataset, the GNN gave $100\%$ node classification accuracy for \texttt{25} of the tested graphs, and an average of $99.98\%$ accuracy across all \texttt{30} graphs. 
For the remaining \texttt{5} graphs, only \texttt{6} nodes, were misclassified and then rectified during post-processing. 
For example, two \textit{design} nodes feeding the XOR gate connecting the Anti-SAT block with the rest of the circuitry got misclassified as part of the Anti-SAT block. 
However, having no KIs in their fan-in, the prediction is ignored. 
For the ITC-99 dataset, the GNN gave $100\%$ node classification accuracy for \texttt{29} out of \texttt{30} tested graphs. 
Only one node was misclassified in one of the \texttt{b21\_C} graphs locked with $K=128$, which was rectified during post-processing. 
Detailed results are reported in Table~\ref{tab:Anti_SAT_results}.

\begin{table}[tb]
\centering
\caption{\textsc{Summary of Generated Datasets}}
\label{tab:Datasets}
\resizebox{\textwidth}{!}{
\setlength\tabcolsep{1.9pt} 
\renewcommand\arraystretch{0.9}
\begin{tabular}{ccccccc}
\hline
\textbf{Dataset} & \textbf{Benchmarks} & \textbf{Circuit Format} & \textbf{\#Classes} & \textbf{$|\hat{f}|$} & \textbf{\#Nodes} & \textbf{\#Circuits} \\ \hline
\multirow{2}{*}{\textbf{Anti-SAT}} & ISCAS-85 & Bench & $2$ & $13$ & $69,468$ & $30$ \\ \cline{2-7} 
 & ITC-99 & Bench & $2$ & $13$ & $450,359$ & $36$ \\ \hline
\multirow{2}{*}{\textbf{TTLock}} & ISCAS-85 & Verilog netlist $65nm$ & $3$ & $34$ & $29,745$ & $45$ \\ \cline{2-7} 
 & ITC-99 & Verilog netlist $65nm$ & $3$ & $34$ & $347,380$ & $54$ \\ \hline
\multirow{3}{*}{\textbf{SFLL-HD$^2$}} & ISCAS-85 & Verilog netlist $65nm$ & $3$ & $34$ & $30,178$ & $45$ \\ \cline{2-7} 
 & \multirow{2}{*}{ITC-99} & Verilog netlist $65nm$ & $3$ & $34$ & $357,374$ & $54$ \\ \cline{3-7} 
 & & Verilog netlist $45nm$ & $3$ & $18$ & $391,411$ & $54$ \\ \hline
\textbf{SFLL-HD$^4$} & ITC-99 & Verilog netlist $65nm$ & $3$ & $34$ & $356,420$ & $54$ \\ \hline
\textbf{SFLL-HD$^{16}$} & ISCAS-85 & Verilog netlist $65nm$ & $3$ & $34$ & $33,354$ & $48$ \\ \hline
\textbf{SFLL-HD$^{32}$} & \multirow{2}{*}{ITC-99} & Verilog netlist $65nm$ & $3$ & $34$ & $465,032$ & $72$ \\ \cline{1-1} \cline{3-7} 
\textbf{SFLL-HD$^{64}$} & & Verilog netlist $65nm$ & $3$ & $34$ & $483,777$ & $72$ \\ \hline
\end{tabular}%
}
\end{table}

\begin{table}[tb]
\centering
\smallerspacecaption
\caption{\textsc{Results of GNNUnlock on Anti-SAT. 
A \textit{design} node and an \textit{Anti-SAT} node are indicated by DN and AN, respectively. 
MN stands for misclassified nodes}}
\label{tab:Anti_SAT_results}
\resizebox{\textwidth}{!}{%
\setlength\tabcolsep{1.9pt} 
\renewcommand\arraystretch{0.8}
\begin{tabular}{ccccccccccc}
\hline
\multirow{2}{*}{\textbf{Test}} & \multirow{2}{*}{\textbf{\begin{tabular}[c]{@{}c@{}}\#Test\\ Graphs\end{tabular}}}& \multirow{2}{*}{\textbf{\begin{tabular}[c]{@{}c@{}}GNN\\ Acc. (\%)\end{tabular}}}& \multicolumn{2}{c}{\textbf{Prec. (\%)}} & \multicolumn{2}{c}{\textbf{Rec. (\%)}} & \multicolumn{2}{c}{\textbf{F1-Score (\%)}} & \multirow{2}{*}{\textbf{\#MN}} & \multirow{2}{*}{\textbf{\begin{tabular}[c]{@{}c@{}}Removal\\ Success (\%)\end{tabular}}} \\ \cline{4-9}
 & & & \textbf{AN} & \textbf{DN} & \textbf{AN} & \textbf{DN} & \textbf{AN} & \textbf{DN} & & \\ \hline
\textbf{c2670}&$8$ &$99.98$ &$99.79$ & $100$ & $100$ &$99.98$ &$99.90$ &$99.99$ & 
\begin{tabular}[c]{@{}c@{}}$2$ DN \end{tabular}& $100$ \\ \hline

\textbf{c3540}&$6$ &$99.98$ &$99.55$ &$99.99$ &$99.78$ &$99.98$ &$99.67$ &$99.99$ & \begin{tabular}[c]{@{}c@{}}$2$ DN\\$1$ AN\end{tabular} & $100$ \\ \hline
\textbf{c5315}&$8$ &$99.99$ &$99.90$ & $100$ & $100$ &$99.99$ &$99.95$ &$99.99$ & $1$ DN & $100$ \\ \hline
\textbf{c7552}&$8$ & $100$ & $100$ & $100$ & $100$ & $100$ & $100$ & $100$ & $-$ & $100$ \\ \hline
\textbf{b14\_C}&$6$ & $100$ & $100$ & $100$ & $100$ & $100$ & $100$ & $100$ & $-$ & $100$ \\ \hline
\textbf{b15\_C}&$6$ & $100$ & $100$ & $100$ & $100$ & $100$ & $100$ & $100$ & $-$ & $100$ \\ \hline
\textbf{b20\_C}&$6$ & $100$ & $100$ & $100$ & $100$ & $100$ & $100$ & $100$ & $-$ & $100$ \\ \hline
\textbf{b21\_C}&$6$ &$99.99$ & $100$ &$99.99$ &$99.92$ & $100$ &$99.96$ &$99.99$ & $1$ AN & $100$ \\ \hline

\textbf{b17\_C}&$6$ & $100$ & $100$ & $100$ & $100$ & $100$ & $100$ & $100$ & $-$ & $100$ \\ \hline
\end{tabular}%
}
\end{table}

\begin{table*}[tb]
\centering
\smallerspacecaption
\caption{\textsc{Results of GNNUnlock on SFLL-HD$^2$. 
The \textit{design}, \textit{perturb}, and \textit{restore} nodes are indicated by DN, PN, and RN, respectively}}
\label{tab:TTLOCK_SFLL_HD_results}
\scriptsize
\resizebox{0.97\textwidth}{!}{%
\renewcommand\arraystretch{0.85}
\begin{tabular}{ccccccccccccccc}
\hline
\multirow{2}{*}{\textbf{Test Set}} & \multirow{2}{*}{\textbf{\begin{tabular}[c]{@{}c@{}}\#Test\\ Graphs\end{tabular}}}& \multirow{2}{*}{\textbf{\begin{tabular}[c]{@{}c@{}}GNN\\ Acc. (\%)\end{tabular}}} & \multicolumn{3}{c}{\textbf{Prec. (\%)}} & \multicolumn{3}{c}{\textbf{Rec. (\%)}} & \multicolumn{3}{c}{\textbf{F1-Score (\%)}} & \multirow{2}{*}{\textbf{\begin{tabular}[c]{@{}c@{}}\#Misclassified\\ Nodes\end{tabular}}} & \multirow{2}{*}{\textbf{\begin{tabular}[c]{@{}c@{}}Removal\\ Success (\%)\end{tabular}}} \\ \cline{4-12}
 & & &\textbf{RN} & \textbf{PN} & \textbf{DN} & \textbf{RN} & \textbf{PN} & \textbf{DN} & \textbf{RN} & \textbf{PN} & \textbf{DN} & & \\ \hline
\textbf{c2670} &$12$&$99.53$ &$99.55$ & $96.36$ & $100$ & $100$ &$99.37$ &$99.46$ &$99.77$ & $97.84$ &$99.73$ & \begin{tabular}[c]{@{}c@{}}$24$ DN as PN\\ $4$ PN as RN\end{tabular} & $100$ \\ \hline
\textbf{c3540}&$9$ &$99.79$ & $98.31$ & $98.97$ & $100$ & $100$ & $98.30$ &$99.88$ &$99.15$ & $98.63$ &$99.94$ & \begin{tabular}[c]{@{}c@{}}$5$ PN as RN\\ $3$ DN as PN\\ $2$ DN as RN\end{tabular} & $100$ \\ \hline
\textbf{c5315 } &$12$ & $100$ & $100$ & $100$ & $100$ & $100$ & $100$ & $100$ & $100$ & $100$ & $100$ &$-$& $100$ \\ \hline
\textbf{c7552 } &$12$ &$99.99$ & $100$ & $100$ &$99.99$ & $100$ &$99.85$ & $100$ & $100$ &$99.93$ &$99.99$ & $1$ PN as DN & $100$ \\ \hline
\textbf{b14\_C } &$9$ &$99.97$ &$99.88$ & $100$ &$99.98$ & $100$ &$99.43$ & $100$ &$99.94$ &$99.72$ &$99.99$ & \begin{tabular}[c]{@{}c@{}}$2$ PN as RN\\ $5$ PN as DN\end{tabular} & $100$ \\ \hline
\textbf{b15\_C } &$9$ &$99.99$ &$99.94$ & $100$ &$99.99$ & $100$ &$99.68$ & $100$ &$99.97$ &$99.84$ &$99.99$ & \begin{tabular}[c]{@{}c@{}}$3$ PN as DN\\ $1$ PN as RN\end{tabular} & $100$ \\ \hline
\textbf{b20\_C } &$9$ &$99.98$ &$99.46$ & $100$ &$99.99$ & $100$ &$99.04$ & $100$ & \multicolumn{1}{l}{$99.73\%$} & \multicolumn{1}{l}{$99.52\%$} & \multicolumn{1}{l}{$99.99\%$} & \begin{tabular}[c]{@{}c@{}}$9$ PN as RN\\ $2$ PN as DN\end{tabular} & $100$ \\ \hline
\textbf{b21\_C } &$9$ & $100$ & $100$ & $100$ & $100$ & $100$ & $100$ & $100$ & $100$ & $100$ & $100$ &$-$& $100$ \\ \hline
\textbf{b22\_C } &$9$ &$99.96$ &$99.94$ & $97.83$ &$99.99$ & $100$ &$99.67$ &$99.96$ &$99.97$ & $98.74$ &$99.98$ & \begin{tabular}[c]{@{}c@{}}$1$ PN as RN\\ $3$ PN as DN\\ $27$ DN as PN\end{tabular} & $100$ \\ \hline
\textbf{b17\_C } &$9$ &$99.94$ &$99.46$ & $95.60$ & $100$ & $100$ &$99.52$ &$99.95$ &$99.73$ & $97.52$ &$99.98$ & \begin{tabular}[c]{@{}c@{}}$3$ DN as RN\\ $57$ DN as PN\\ $6$ PN as RN\end{tabular} & $100$ \\ \hline
\end{tabular}%
}
\end{table*}

Having $100\%$ node classification accuracy for a given benchmark locked with different key-sizes (such as the case for \texttt{c7552}, \texttt{b14\_C}, etc.) shows that \textbf{the ratio of sizes between Anti-SAT block and design does not affect the performance of GNNUnlock and overall recovery of the original design}. 
Obtaining $100\%$ node classification accuracy for Anti-SAT is validated since the Anti-SAT block has a specific structure controlled by external KIs, making it easy to be learned.

\subsection{Breaking SFLL-HD$^{h>0}$ and TTLock with GNNUnlock}
\label{sec:breaking_ttlock}

For the SFLL-HD$^2$ ISCAS-85 dataset, the GNN gave $100\%$ node classification accuracy for \texttt{20} tested graphs and an average of $99.83\%$ accuracy across all \texttt{45} ISCAS-85 graphs. 
In total, \texttt{38} nodes were misclassified then rectified using post-processing. 
For example, \texttt{27} \textit{design} nodes, coming from the different \texttt{25} graphs were misclassified as \textit{perturb} nodes. 
\textit{The misclassified nodes belonged to NOR-tree-like structures in the original designs.}
This misclassification did not affect the removal of the protection logic because non-protected input patterns controlled these nodes and hence predictions were dropped.
For SFLL-HD$^2$ ITC-99 dataset, the GNN gave $100\%$ node classification accuracy for \texttt{29} of the tested graphs and an average of $99.97\%$ accuracy across all \texttt{54} graphs (Table~\ref{tab:TTLOCK_SFLL_HD_results}).

For the TTLock ISCAS-85 dataset, the GNN gave $100\%$ node classification accuracy for $29$ of the tested graphs and an average of $99.94\%$ accuracy across all $45$ graphs. 
For the remaining $16$ graphs, only $19$ nodes were misclassified. 
The results show that \textbf{GNNUnlock can accurately detect all three types of nodes} with precision and recall values ranging from $93.17\%$ to $100\%$. 
For the TTLock ITC-99 dataset, the GNN gave $100\%$ node classification accuracy for $16$ of the tested graphs and an average of $99.95\%$ accuracy across all $54$ graphs. 
Comparing accuracy results on ISCAS-85 Vs. ITC-99 (for both SFLL-HD$^{h>0}$ and TTLock) confirms that \textbf{GNNUnlock has a consistent performance for predicting the protection logic regardless of the benchmark size}.
Table~\ref{tab:results_hd4_nan} represents all the averaged results for TTLock datasets.
It was observed that the \textit{restore} predictor is $100\%$ successful (with $100\%$ precision and recall) in all ITC-99 test cases. This also confirms that the separation between the \textit{perturb} and \textit{design} nodes is challenging.

\textit{To study the effect of $h$ on the performance of GNNUnlock}, we launch the attack on the SFLL-HD$^4$ ITC-99 dataset. 
Note: due to lack of space, we only report the averaged metrics in Table~\ref{tab:results_hd4_nan}. 
Here, we also report the results of GNNUnlock on the SFLL-HD$^2$ ITC-99 dataset with a technology node of $45nm$ to evaluate the effect of the target library on the attack's performance. 
The trend is the same in all cases, while there is a small difference in the values. 
GNNUnlock detects all three types of nodes accurately with average accuracy values of $99.94\%$, $99.97\%$, and $99.90\%$, for SFLL-HD$^2$ in $45nm$, SFLL-HD$^2$ in $65nm$, and SFLL-HD$^4$ in $65nm$ datasets, respectively. 
After post-processing, all nodes were classified correctly ($100\%$), verifying that \textbf{GNNUnlock can handle different technology nodes and various parameter settings.}

\subsection{Comparison with State-of-the-art Attacks}
\label{sec:corner_cases}

To demonstrate the superiority of GNNUnlock, a set of ITC-99 benchmarks are locked using SFLL-HD with $K/h=2$, generating two datasets: one for $K=128$, $h=64$ and one for $K=64$, $h=32$.
The ISCAS-85 benchmarks are locked using $K=32$, $h=16$. 
The characteristics are listed in Table~\ref{tab:Datasets}. When the FALL attacks~\cite{sirone2020functional} were launched on these benchmarks, they reported $0$ keys. Next, the attack platform of SFLL-HD-Unlocked~\cite{yang2019stripped} was launched on the same benchmarks, which failed to identify the perturb signals and did not recover the secret key. 
However, our GNNUnlock was $100\%$ successful in retrieving the unlocked designs in all of the cases.
The results are documented in Table~\ref{tab:results_hd4_nan}.

\begin{table}[tb]
\centering
\caption{\textsc{Examining the effect of $h$ value and technology node on the performance of GNNUnlock. TR stands for training}}
\label{tab:results_hd4_nan}
\resizebox{\textwidth}{!}{%
\setlength\tabcolsep{1.9pt}
\begin{tabular}{ccccccccc}
\hline
\textbf{Dataset} &\textbf{Benchmarks} &\textbf{\begin{tabular}[c]{@{}c@{}}Tech.\\ Node (nm)\end{tabular}} & \textbf{\begin{tabular}[c]{@{}c@{}}GNN\\ Acc. (\%)\end{tabular}} & \textbf{\begin{tabular}[c]{@{}c@{}}Avg.\\ Prec. (\%)\end{tabular}} & \textbf{\begin{tabular}[c]{@{}c@{}}Avg.\\ Rec. (\%)\end{tabular}} & \textbf{\begin{tabular}[c]{@{}c@{}}Avg.\\ F1-Score (\%)\end{tabular}} & \textbf{\begin{tabular}[c]{@{}c@{}}Removal\\ Success (\%)\end{tabular}}& \textbf{\begin{tabular}[c]{@{}c@{}}Avg.\\ TR Time ($s$)\end{tabular}} 
\\ \hline
\textbf{TTLock} & ISCAS-85 &$45$ & $99.94$ &	$99.20$&	$99.53$	&$99.33$& $100$&$1,910$\\ \hline
\textbf{TTLock} & ITC-99 &$45$ & $99.95$ &	$97.48$&	$99.36$&	$98.33$ & $100$&$22,777$\\ \hline
\textbf{SFLL-HD$^2$} &ITC-99& $45$ & $99.94$ & $98.73$ &$99.77$ &$99.21$ & $100$&$24,123$ \\ \hline
\textbf{SFLL-HD$^2$} &ITC-99& $65$ & $99.97$ &$99.56$ &$99.85$ &$99.70$ & $100$ &$22,052$\\ \hline
\textbf{SFLL-HD$^4$} &ITC-99& $65$ &$99.90$ & $98.40$ &$99.60$ & $98.95$ & $100$&$233,017$ \\ \hline
\textbf{SFLL-HD$^{16}$} &ISCAS-85& $65$ &$99.24$ & $97.68$ & $97.30$ & $97.40$ & $100$&$1,907$ \\ \hline
\textbf{SFLL-HD$^{32}$} &ITC-99& $65$ &$99.83$ & $97.56$ & $98.45$ & $97.96$ & $100$&$28,049$ \\ \hline
\textbf{SFLL-HD$^{64}$} &ITC-99& $65$ &$99.69$&	$97.60$&	$97.96$	&$97.74$&$100$&$32,494$\\\hline
\end{tabular}
}
\end{table}
\section{Conclusion}
\label{sec:Conclusion}

In this paper, we proposed the first-of-its-kind scheme for unlocking provably secure logic locking (PSLL), leveraging a graph neural network (GNN) that learns the common structural features of the protection logic added by such techniques. 
The GNN does not learn a syntactic implementation but absorbs the protection logic trend, allowing it to handle variants naturally.
We also developed a post-processing algorithm to rectify any misclassifications by the GNN, depending solely on the nodes' connectivity, the GNN predictions, and the known properties of the protection logic. 
We demonstrated the superiority of GNNUnlock by comparing it to two state-of-the-art attacks in unlocking three PSLL techniques, Anti-SAT, TTLock, and SFLL-HD. 
Our GNNUnlock scheme was able to break all locked benchmarks while state-of-the-art attacks struggled with some corner cases.
We believe GNNUnlock opens up new frontiers in advancing the state-of-the-art defenses and attacks in logic locking.

\section*{Acknowledgements}
This work was carried out in part on the High-Performance Computing resources at Khalifa University.
Besides, this work is supported in part by the Center for Cyber Security (CCS) at New York University Abu Dhabi (NYUAD).

\bibliography{main}
\bibliographystyle{IEEEtran} 

\end{document}